\documentclass[prb,twocolumn,showpacs,amsmath,superscriptaddress,psfig,amssymb]{revtex4}

\usepackage{listings}
\usepackage{graphicx}
\usepackage{dcolumn}
\usepackage{bm}

\begin{document}

\title{Upper critical field and thermally activated flux flow in single crystalline Tl$_{0.58}$Rb$_{0.42}$Fe$_{1.72}$Se$_2$}
\author{L. Jiao}\affiliation{Department of Physics, Zhejiang University, Hangzhou, Zhejiang 310027, China}
\author{Y. Kohama}\affiliation{Los Alamos National Laboratory, Los Alamos, NM 87545, USA}
\author{J. L. Zhang}\affiliation{Department of Physics, Zhejiang University, Hangzhou, Zhejiang 310027, China}
\author{H. D. Wang}\affiliation{Department of Physics, Zhejiang University, Hangzhou, Zhejiang 310027, China}
\affiliation{Department of Physics, Hangzhou Normal University,
Hangzhou, Zhejiang 310036, China}
\author{B. Maiorov}\affiliation{Los Alamos National Laboratory, Los Alamos, NM 87545, USA}
\author{F. F. Balakirev}\affiliation{Los Alamos National Laboratory, Los Alamos, NM 87545, USA}
\author{Y. Chen}\affiliation{Department of Physics, Zhejiang University, Hangzhou, Zhejiang 310027, China}
\author{L. N. Wang}\affiliation{Department of Physics, Zhejiang University, Hangzhou, Zhejiang 310027, China}
\author{T. Shang}\affiliation{Department of Physics, Zhejiang University, Hangzhou, Zhejiang 310027, China}
\author{M. H. Fang}\affiliation{Department of Physics, Zhejiang University, Hangzhou, Zhejiang 310027, China}
\author{H. Q. Yuan}\email{hqyuan@zju.edu.cn}\affiliation{Department of Physics, Zhejiang University, Hangzhou, Zhejiang 310027, China}
\date{\today}

\begin{abstract}
The upper critical field $\mu_0H_{c2}(T_c)$ of
Tl$_{0.58}$Rb$_{0.42}$Fe$_{1.72}$Se$_2$ single crystals has been
determined by means of measuring the electrical resistivity in both
a pulsed magnetic field ($\sim$58T) and a DC magnetic field
($\sim$14T). It is found that $\mu_0H_{c2}$ linearly increases with
decreasing temperature for $\textbf{H}$$\parallel$$c$, reaching
$\mu_0H_{c2}^{\textbf{H}\parallel c}(0\textrm{K})\simeq60$ T. On the
other hand, a larger $\mu_0H_{c2}(0\textrm{K})$ with a strong convex
curvature is observed for $\textbf{H}$$\perp$$c$
($\mu_0H_{c2}^{\textbf{H}\perp c}$(18K)$\simeq$60T). This compound
shows a moderate anisotropy of the upper critical field around
$T_c$, which decreases with decreasing temperature. Analysis of the
upper critical field based on the Werthamer-Helfand-Hohenberg (WHH)
method indicates that $\mu_0H_{c2}(0\textrm{K})$ is orbitally
limited for $\textbf{H}$$\parallel$$c$, but the effect of spin
paramagnetism may play an important role on the pair breaking for
$\textbf{H}$$\perp$$c$. All these experimental observations
remarkably resemble those of the iron pnictide superconductors,
suggesting a universal scenario for the iron-based superconductors.
Moreover, the superconducting transition is significantly broadened
upon applying a magnetic field, indicating strong thermal
fluctuation effects in the superconducting state of
Tl$_{0.58}$Rb$_{0.42}$Fe$_{1.72}$Se$_2$. The derived thermal
activation energy for vortex motion is compatible with those of the
1111-type iron pnictides.
\end{abstract}

\maketitle

\section{Introduction}
The newly discovered Fe-based superconductors (FeSCs) share many
similarities to the high $T_c$ cuprates,\cite{1} e.g., both showing
a relatively high superconducting transition temperature $T_c$ and
possessing a layered crystal structure. It is, therefore, natural to
compare these two classes of superconductors, which might help
unravel the puzzles of high $T_c$ superconductivity. However,
significantly distinct properties have been demonstrated in the
FeSCs,\cite{1} including that (i) most of the parent compounds of
FeSCs are typical `bad metal' instead of a Mott insulator as found
in the cuprates; (ii) the FeSCs are a multi-band system, which seems
to favor a $\emph{s}^{\pm}$ pairing state rather than a
\emph{d}-wave state; (iii) both the FeSCs and the cuprates possess a
very large upper critical field, but the FeSCs show nearly isotropic
$H_{c2}$ at low temperatures despite of their layered crystal
structures. Clarification of the electronic coupling strength in
FeSCs is the basis for establishing a pertinent theory of
superconductivity. Various approaches, either based on the Fermi
surface nesting\cite{DJ} or started from the proximity to a
Mott-insulator\cite{QMS} were initially proposed to reveal the
physics of iron pnictides, but no consensus has been reached.
Recently, dual characters of localized and itinerant 3d-electrons
were theoretically proposed\cite{Kou} and experimentally shown in
some iron pnictides.\cite{Yuan1,Moon} To reveal the nature of
magnetism and superconductivity in FeSCs and compare it with the
high $T_c$ cuprates, it remains highly desired to search for FeSCs
nearby a Mott insulator.

Very recently a new class of FeSCs, AFe$_x$Se$_2$ (A=K,\cite{GuoJin}
Cs,\cite{Krzton} Rb,\cite{LiShen} (Tl$_{1-y}$K$_y$)\cite{Fang} and
(Tl$_{1-y}$Rb$_y$)\cite{WHD}), were discovered with $T_c$ up to
$\sim$33K. Remarkably different from the iron pnictides,
superconductivity in iron selenides seems to develop from an
antiferromagnetic Mott insulator with a rather high N\'{e}el
temperature.\cite{Fang,WHD,Shermadini,LiuRH,Bao,CXH1104} In these
compounds, one may tune the interplay of superconductivity and
magnetism by changing the Fe-vacancy order.\cite{Fang,Bao,CXH1104}
Furthermore, the reported ARPES experiments on iron selenides showed
that an isotropic superconducting gap emerges around the electron
pocket at M point but the hole band centered at $\Gamma$ point sinks
below the Fermi level.\cite{FDL,DH1,ZXJ} This is in sharp contrast
to that of the iron pnictide superconductors, in which both hole-
and electron-pockets, connected with a nesting wave vector, were
experimentally observed.\cite{arpes} It is, therefore, of great
interest to find out whether the iron selenide superconductors
represent a new type of FeSCs (e.g., similar to the high $T_c$
cuprates) or remain similar to other iron pnictides. In any case,
the iron selenide superconductors may provide an alternative example
for studying the pairing mechanisms of high $T_c$ superconductivity,
in particular for the FeSCs. To elucidate the above issues, it is
highly important to compare the main superconducting parameters of
the iron selenides with those of the iron pnictides, and also among
the iron selenide series.

In this article, we report measurements of the electrical
resistivity in both a pulsed magnetic field and a DC magnetic field
for the single-crystalline Tl$_{0.58}$Rb$_{0.42}$Fe$_{1.72}$Se$_2$.
It is found that Tl$_{0.58}$Rb$_{0.42}$Fe$_{1.72}$Se$_2$ shows a
very large upper critical field ($\mu_0H_{c2}^{\textbf{H}\parallel
c}(0\textrm{K})\simeq60$T, $\mu_0H_{c2}^{\textbf{H}\perp
c}(18\rm{K})$$\simeq60$T) with a moderate anisotropic parameter
$\gamma$ ($\gamma=H_{c2}^{\textbf{H}\perp
c}/H_{c2}^{\textbf{H}\parallel c}$), remarkably resembling those of
iron pnictide
superconductors.\cite{Yuan,ref2,MHF11,ZJL111,Baily,Kano} On the
other hand, the superconducting transition of
Tl$_{0.58}$Rb$_{0.42}$Fe$_{1.72}$Se$_2$ is substantially broadened
in a magnetic field, indicating significant contributions of
thermally activated flux flow in the vortex state.

\section{Experimental methods}

Single crystals of Tl$_{0.58}$Rb$_{0.42}$Fe$_{1.72}$Se$_2$ were
synthesized by using a Bridgeman method.\cite{WHD} The X-ray
diffraction (XRD) identified the derived samples as a single phase
with a tetragonal ThCr$_2$Si$_2$ crystal structure. The actual
composition of the crystals was determined by energy dispersive
X-ray spectrometer (EDXS). Magnetic field dependence of the
electrical resistivity, $\rho(H)$, was measured up to 57.5T using a
typical 4-probe method in a capacitor-bank-driven pulsed magnet. The
experimental data were recorded on a digitizer using a custom
designed high-resolution, low-noise synchronous lock-in technique.
In order to minimize the eddy-current heating caused by the pulsed
magnetic field, very small crystals were cleaved off along the
\emph{ab}-plane from the as-grown samples. The electrical
resistivity in a DC magnetic field (0-14T) was measured in an Oxford
Instruments HELIOX VL system using a Lakeshore AC Resistance Bridge
and the angular dependence of the electrical resistivity was
performed in a Quantum Design (QD) Physical Properties Measurement
System (9T PPMS). Angular linear ($\rho$) transport measurements
were carried out using the maximum Lorentz force configuration
(\textbf{J}$\perp$\textbf{H}), with \textbf{H} applied at an angle
$\theta$ from the \emph{c} axis of the crystal (see the inset of
Fig. \ref{figangle}).

\section{The upper critical field and its anisotropy}

 \begin{figure}
 \includegraphics[width=0.45\textwidth]{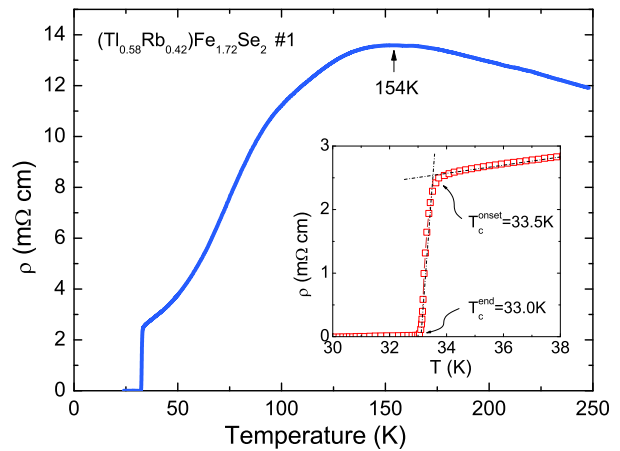}
\caption{(Color online) Temperature dependence of the electrical
resistivity $\rho(T)$ at $\mu_0H=0$ for
Tl$_{0.58}$Rb$_{0.42}$Fe$_{1.72}$Se$_2$(\#1). The inset enlarges the
section at the superconducting transition, where
$T_c^{onset}\simeq33.5$K and $T_c^{end}\simeq33$K are determined
from the onset and the end-point of the superconducting transition,
respectively.} \label{fig1}
\end{figure}

In Fig.~\ref{fig1}, we show the temperature dependence of the
electrical resistivity at zero field for
Tl$_{0.58}$Rb$_{0.42}$Fe$_{1.72}$Se$_2$ (\#1). One can see that the
resistivity $\rho(T)$ shows a hump around 154K, changing from
semiconducting to metallic behavior upon cooling down from room
temperature. Such a hump in $\rho(T)$ has been widely observed in
the iron selenides\cite{GuoJin,Krzton,LiShen,Fang,WHD,LiuRH,Guojing}
and its position can be tuned either by doping\cite{LiuRH} or
pressure.\cite{Guojing} The origin of the hump and its relation with
superconductivity remain unclear. A very sharp superconducting
transition shows up at $T_c\simeq$33.5K, indicating a high quality
of the sample. Note that we have totally measured four samples cut
from the same batch in this context and their $T_c$ only varies
slightly from 32.9K to 33.5K, indicating a good reproducibility of
the superconducting properties in these samples.

\begin{figure}
\includegraphics[width=0.45\textwidth]{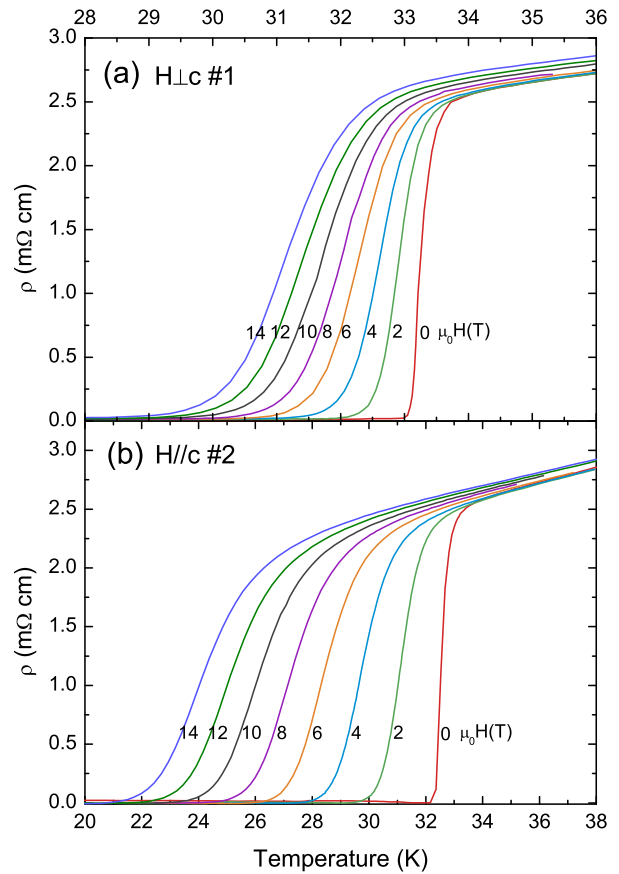}
\caption{(Color online) Temperature dependence of the electrical
resistivity $\rho(T)$ for
Tl$_{0.58}$Rb$_{0.42}$Fe$_{1.72}$Se$_2$(\#1 and \#2) measured in DC
fields up to 14T: (a) $\textbf{H}$$\perp$$c$; (b)
$\textbf{H}$$\parallel$$c$.} \label{fig2DC}
\end{figure}

\begin{figure}
\includegraphics[width=0.45\textwidth]{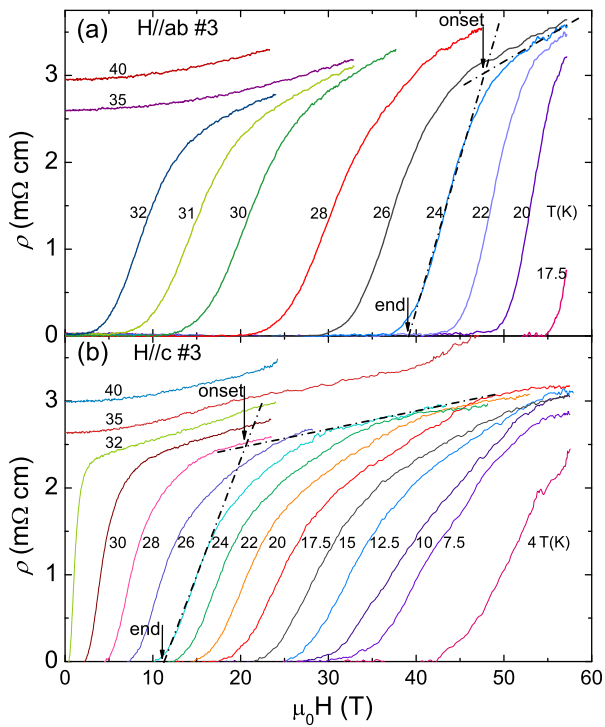}
\caption{(Color online) Magnetic field dependence of the electrical
resistivity $\rho(\mu_0H)$ at various temperatures for
Tl$_{0.58}$Rb$_{0.42}$Fe$_{1.72}$Se$_2$(\#3): (a)
$\textbf{H}$$\perp$$c$; (b) $\textbf{H}$$\parallel$$c$.}
\label{fig2}
\end{figure}

Fig. \ref{fig2DC} and Fig. \ref{fig2} show the temperature- and the
magnetic-field-dependence of the electrical resistivity for
Tl$_{0.58}$Rb$_{0.42}$Fe$_{1.72}$Se$_2$, respectively. In order to
study the anisotropic behavior, the electrical resistivity was
measured with field perpendicular (a) and parallel (b) to the
\emph{c}-axis. The magnetic field is applied up to 14T for the DC
fields (Fig. \ref{fig2DC}) and up to 58T for the pulsed magnetic
fields (Fig. \ref{fig2}). Obviously, the superconducting transition
eventually shifts to lower temperature upon applying a magnetic
field. However, superconductivity is remarkably robust against the
magnetic field in Tl$_{0.58}$Rb$_{0.42}$Fe$_{1.72}$Se$_2$ and it is
not yet completely suppressed at our maximum field of 58T.
Furthermore, one can see that the superconducting transition is
significantly broadened upon applying a magnetic field, showing a
tail structure at low temperature. For example, the width of the
superconducting transition, defined from the onset temperature to
the end point of the superconducting transition (see the inset of
Fig.1), is as small as 0.5K at zero field but increases to 2K and
3.6K for $\textbf{H}$$\perp$$c$ and $\textbf{H}$$\parallel$$c$ at
14T, respectively. Similar features were also observed in some
1111-type iron pnictides.\cite{Jaroszynski,Shahbazi,SmFeAsO} We will
argue later that such behavior might be attributed to the thermally
activated flux flow in the vortex state. In the normal state,
Tl$_{0.58}$Rb$_{0.42}$Fe$_{1.72}$Se$_2$ shows significant positive
magnetoresistance for both $\textbf{H}$$\parallel$$c$ and
$\textbf{H}$$\perp$$c$. It is noted that, in iron pnictides, the
magnetoresistance becomes very large while entering the magnetic
state, but is negligible in the non-magnetic state.\cite{Yuan1} One
possibility for the occurrence of such a large magnetoresistance in
Tl$_{0.58}$Rb$_{0.42}$Fe$_{1.72}$Se$_2$ might be related to its
magnetic ordering at high temperature.\cite{Bao,CXH1104}

 \begin{figure}
 \includegraphics[width=0.45\textwidth]{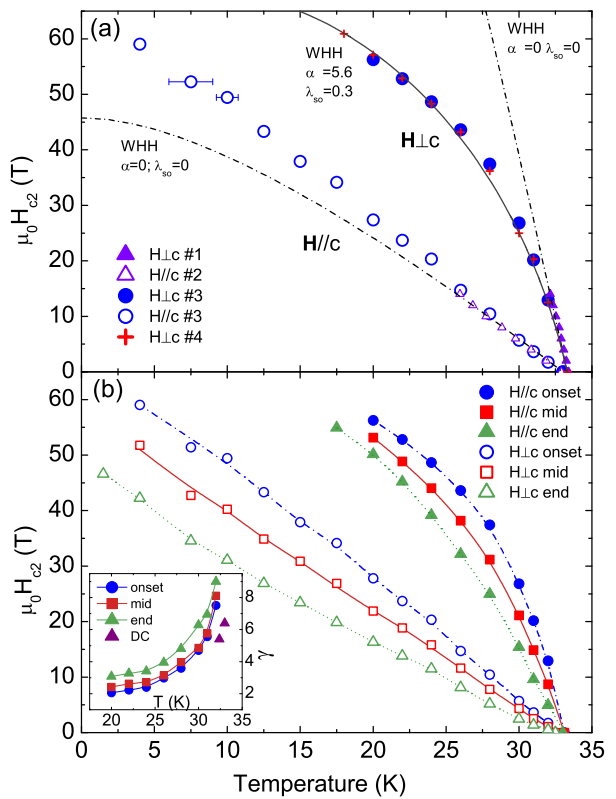}
\caption{(Color online) The upper critical field $\mu_0H_{c2}(T_c)$
for Tl$_{0.58}$Rb$_{0.42}$Fe$_{1.72}$Se$_2$. (a) The values of
$\mu_0H_{c2}(T_c)$ are determined from the superconducting onsets as
described in Fig. 1 and Fig. 3. Symbols of the open circle
($\circ$), filled circle ($\bullet$) and cross (+) represent the
data obtained in a pulsed magnetic field and the triangles
($\triangle$ and $\blacktriangle$) denote those measured in a DC
magnetic field. Note that samples \#3 and \#4 were measured in a
pulsed field, but only sample \#3 was successfully measured for both
$\textbf{H}$$\perp$$c$ and $\textbf{H}$$\parallel$$c$. (b) The
values of $\mu_0H_{c2}(T_c)$ are determined from the onset
($\circ$), the min-point ($\Box$) and the end-point ($\triangle$) of
the resistive superconducting transitions (sample \# 3),
respectively. The inset shows the corresponding temperature
dependence of the anisotropic parameter $\gamma(T)$.} \label{fig3}
\end{figure}

The upper critical field $\mu_0H_{c2}(T_c)$  of
Tl$_{0.58}$Rb$_{0.42}$Fe$_{1.72}$Se$_2$ is shown in Fig.~\ref{fig3},
in which various symbols represent either different field
orientations for the same sample (\#3) or different samples as
marked in the figure. In Fig. 4(a), we determine the critical
temperatures $T_{c2}$ (in the case of DC field) or the critical
fields $\mu_0H_{c2}$ (in the case of pulsed field) from the
superconducting onsets as described in the inset of Fig. 1 and also
in Fig.3, i.e., the intersection point of the resistive curves in
the normal state and the superconducting transition. Such a
determination of $\mu_0H_{c2}$ (or $T_{c2}$) is appropriate for the
in-field measurements and is particularly useful in presence of the
magnetoresistance $\rho(\mu_0H)$. In this case, one can extrapolate
$\rho(\mu_0H)$ to lower temperatures to determine $\mu_0H_{c2}$ at
the lowest temperatures since $\rho(\mu_0H)$ in the normal state
hardly depend on temperature. It is noted that similar field-induced
broadening of the resistive superconducting transition was also
observed in the high ${T_c}$ cuprates in which the onset temperature
as we described here was shown to be close to that determined by
other bulk measurements, e.g., the magnetization.\cite{Welp1} From
Fig. 4(a), one can see that the derived $\mu_0H_{c2}(T_c)$
demonstrates quantitatively same behavior for all the investigated
samples, independent of the detailed experimental methods. The upper
critical field $\mu_0H_{c2}^{\textbf{H}\parallel c}(T_c)$ linearly
increases with decreasing temperature, reaching a value of
$\mu_0H_{c2}^{\textbf{H}\parallel c}(0\textrm{K})\simeq60$T. On the
other hand, $\mu_0H_{c2}^{\textbf{H}\perp c}(T_c)$ shows a convex
curvature with a much larger value at low temperatures. Such
behavior of $\mu_0H_{c2}(T_c)$ is not changed by the field-induced
broadening of the superconducting transition. Taking sample \#3 as
an example, in Fig. 4(b) we plot the upper critical fields
$\mu_0H_{c2}(T_c)$ determined at the onset, mid- and end-point of
the resistive transitions (see Fig. 3), which follow remarkably
similar temperature dependence. We note that our results are
consistent with those of K$_{0.8}$Fe$_{1.76}$Se$_2$\cite{Mun} and
K$_{0.76}$Fe$_{1.61}$Se$_{0.96}$S$_{1.04}$;\cite{LHC} the former was
measured using a tunnel-diode resonator (TDR) technique in a pulsed
magnetic field and the latter was measured  only up to 9T. It was
also shown that the values of $\mu_0H_{c2}$ determined from the
end-point of the resistive transitions and the TDR technique are
consistent, \cite{Mun} further confirming the validity of our
methods in the determination of upper critical fields.

In a superconductor, the Cooper pairs can be destroyed by the
following two mechanisms in a magnetic field: (i) the orbital pair
breaking due to the Lorentz force acting via the charge on the
momenta of the paired electrons (orbital limit); (ii) the Zeeman
effect aligning the spins of the two electrons with the applied
field (Pauli paramagnetic limit). According to the WHH method, the
orbital-limiting upper critical field $\mu_0H_{c2}^{orb}$(0K) for a
single band BCS superconductor is determined by the initial slope of
$\mu_0H_{c2}(T_c)$ at $T_c$, i.e.\cite{WHH}
\begin{equation}
\mu_0H_{c2}^{orb}(0\textrm{K})=-0.69\it{T_c(dH_{c2}/dT)\mid_{T=T_c}},
\label{eq:1}
\end{equation}
which value may depend on the field orientations. The Pauli
paramagnetic limiting field for weakly coupled BCS superconductors
is given by\cite{Clo}
\begin{equation}
\mu_0H_{c2}^p(0\textrm{K})[\textrm{T}]=1.86\it{T_c}\rm[K].\label{eq:2}
\end{equation}

While the upper critical field is usually restricted by the orbital
limit in conventional superconductors, the spin paramagnetic effect
may play an important role in pair breaking in some unconventional
superconductors. Tl$_{0.58}$Rb$_{0.42}$Fe$_{1.72}$Se$_2$ reveals a
relatively large and anisotropic initial slope of $\mu_0H_{c2}(T_c)$
near $T_c$, which reaches a value of -12T/K and -2T/K (from the
superconducting onsets) for $\textbf{H}$$\perp$$c$ and
$\textbf{H}$$\parallel$$c$, respectively. Following Eq.~\ref{eq:1},
one can derive the orbitally limited upper critical field, which
gives $\mu_0H_{c2}^{orb}(0\textrm{K})=273$T for
$\textbf{H}$$\perp$$c$ and 45T for $\textbf{H}$$\parallel$$c$. As
shown in Fig.~\ref{fig3}, $\mu_0H_{c2}^{orb}(0\textrm{K})$
considerably exceeds the experimental value of
$\mu_0H_{c2}(0\textrm{K})$ for $\textbf{H}$$\perp$$c$, but lightly
falls below the corresponding $\mu_0H_{c2}(0\textrm{K})$ for
$\textbf{H}$$\parallel$$c$. On the other hand, Eq.~\ref{eq:2} gives
a Pauli paramagnetic limiting field of
$\mu_0H_{c2}^p(0\textrm{K})\approx60$T in terms of the BCS theory.
Thus, it is likely that the upper critical field is limited by
orbital effect for $\textbf{H}$$\parallel$$c$, but by spin
paramagnetic effect for $\textbf{H}$$\perp$$c$. In order to further
look into this point, we fitted the experimental data of
$\mu_0H_{c2}(T_c)$ by the WHH model,\cite{WHH} in which the effects
of both orbital- and spin-pair breaking are considered (see Fig.
4a). In this model, $\alpha$ and $\lambda_{so}$ are the fitting
parameters; $\alpha$ is the Maki parameter, which represents the
relative strength of spin and orbital pair-breaking, and
$\lambda_{so}$ is the spin-orbit scattering constant. As shown in
Fig.4(a), the upper critical field $\mu_0H_{c2}(T_c)$ for both
$\textbf{H}$$\parallel$$c$ and $\textbf{H}$$\perp$$c$ are not well
described by the WHH method while ignoring the spin effect (see the
dot dashed lines with $\alpha$=0 and $\lambda_{so}$=0). The
enhancement of $\mu_0H_{c2}^{\textbf{H}\parallel c}(T_c)$ at low
temperature is likely attributed to its multi-band electronic
structure as discussed in other FeSCs.\cite{Kano} Indeed,
$\mu_0H_{c2}^{\textbf{H}\perp c}(T_c)$ can be well fitted by the WHH
model after considering the spin effect (see the solid line in
Fig.4a), which gives $\alpha$=5.6 and $\lambda_{so}$=0.3. Such a
large value of $\alpha$ indicates that the spin paramagnetism may
play an important role on suppressing superconductivity for
$\textbf{H}$$\perp$$c$. Similar analysis applies to the data of
$\mu_0H_{c2}(T_c)$ derived at various resistive drops of the broad
superconducting transition, showing generally consistent behavior.
In the case of a cylinder-like Fermi surface, the open electron
orbits along the \emph{c}-axis make the orbital limiting upper
critical field unlikely. Existence of a Pauli limiting
$\mu_0H_{c2}(0)$ for $H\perp c$ seems to agree with the enhanced
anisotropy in Tl$_{0.58}$Rb$_{0.42}$Fe$_{1.72}$Se$_2$ (see below).
On the other hand, its multi-band electronic structure may
complicate the analysis of $\mu_0H_{c2}(T_c)$. Nevertheless, the
upper critical field of Tl$_{0.58}$Rb$_{0.42}$Fe$_{1.72}$Se$_2$
shows similar behavior to that of other iron pnictide
superconductors,\cite{Yuan,ref2,MHF11,ZJL111,Baily,Kano} indicating
a uniform scenario of $\mu_0H_{c2}(0)$ in the FeSCs.

The anisotropic parameter $\gamma(T)$ of
Tl$_{0.58}$Rb$_{0.42}$Fe$_{1.72}$Se$_2$ (sample \#3), derived at
variant points of the superconducting transition, is shown in the
inset of Fig.4(b). One can see that all these curves follow exactly
same temperature dependence but with a small deviation on their
absolute values of $\gamma (T)$. For example, the anisotropic
parameter $\gamma$, determined from the superconducting onsets, is
as high as 8 near $T_c$ but reaches 2 at $T=20$ K. Similar values of
$\gamma(T)$ were also obtained for other samples as a result of
their consistent behavior of the upper critical field (see Fig. 4a).
The anisotropy of the upper critical field near $T_c$ can be
associated with its electronic band structure. Observation of
$\gamma \sim 8$ near $T_c$ might suggest a quasi-2D electronic
structure for Tl$_{0.58}$Rb$_{0.42}$Fe$_{1.72}$Se$_2$, which is
compatible with the large resistive anisotropy in the normal state
($\rho_c/\rho_{ab}\sim 30$)\cite{WHD}. An anisotropy of
$\gamma$$\sim$8 near $T_c$ in
Tl$_{0.58}$Rb$_{0.42}$Fe$_{1.72}$Se$_2$ is relatively large among
the FeSCs, \cite{Yuan,ref2,MHF11,ZJL111,Baily,Kano} but it is close
to that of the 1111-compounds. \cite{Baily} Nevertheless, in all
these systems the superconducting properties tend to be more
isotropic at low temperatures, which would be compatible with the
isotropic-like superconducting energy gaps observed in the ARPES
experiments around 15K. \cite{FDL,DH1,ZXJ,arpes}

\begin{figure}
 \includegraphics[width=0.45\textwidth]{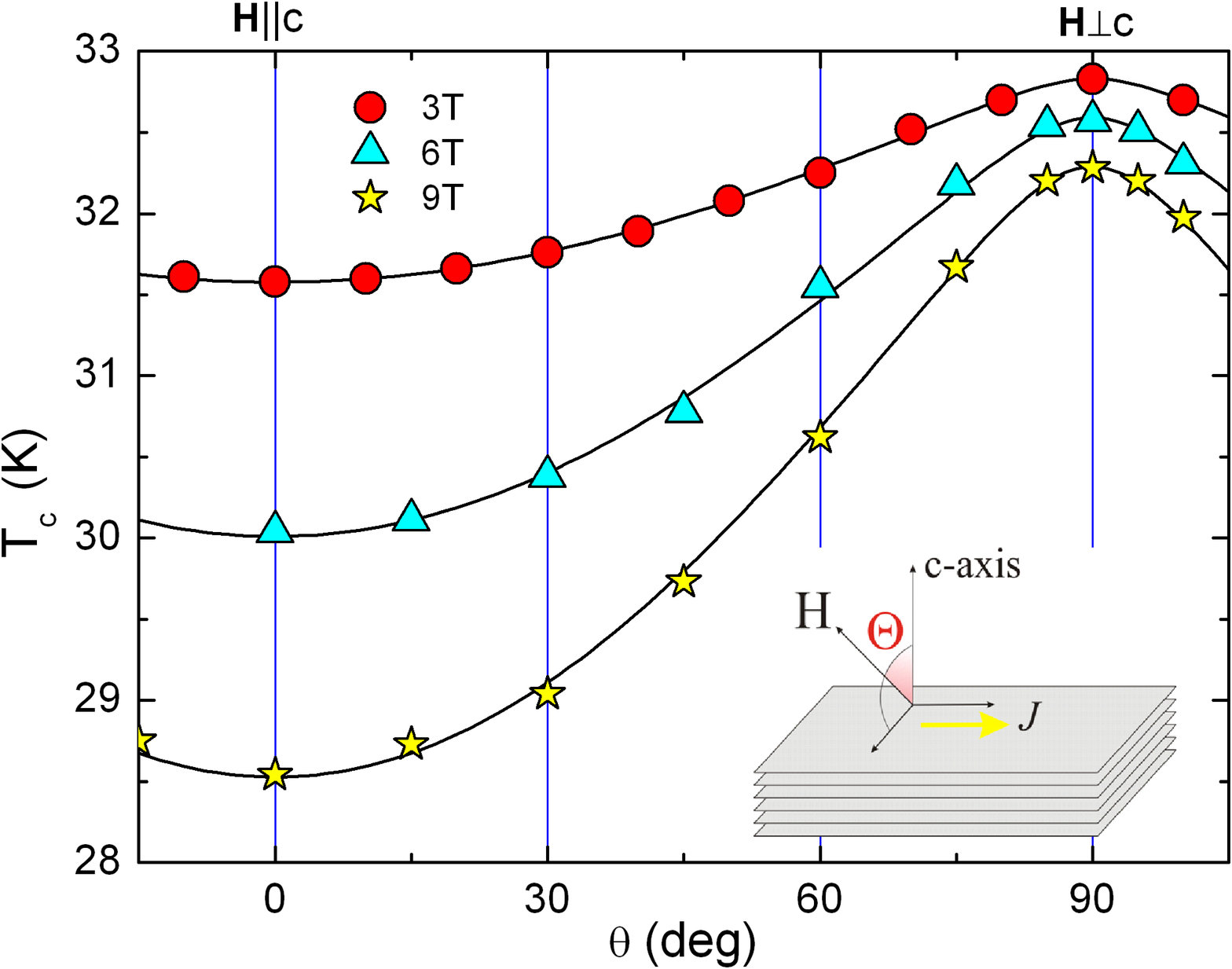}
\caption{(Color online) The angular dependence of the
superconducting critical temperature $T_c(\theta)$ at magnetic
fields of $\mu_0H$=3T, 6T and 9T. The solid lines show the fittings
to Eq.~\ref{eq:4}. Here $T_c(\theta)$ is determined from the
superconducting onsets.} \label{figangle}
\end{figure}

In order to further characterize the nature of the anisotropy in
Tl$_{0.58}$Rb$_{0.42}$Fe$_{1.72}$Se$_2$, we have measured the
angular dependence of the electrical resistivity $\rho(T)$ at
various magnetic fields. Fig. \ref{figangle} plots the angular
dependence of $T_c(\theta)$, where $\theta$ is the angle between the
magnetic field and the \emph{c}-axis of the sample as marked in the
figure.

According to the single band anisotropic Ginzburg-Landau (G-L)
theory,\cite{Tilley,GBlatter} the angular dependence of the upper
critical field can be scaled by:
\begin{equation}
\mu_0H_{c2}^{GL}(\theta)=\mu_0H_{c2}/\sqrt{cos^2(\theta)+\gamma^{-2}sin^2(\theta)},
\label{eq:3}
\end{equation}
where $\gamma$=($m_{ab}/m_c$)$^{1/2}$=$H_{c2}^{\textbf{H}\perp
c}/H_{c2}^{\textbf{H}\parallel c}$. Here $m_{ab}$ and $m_c$ are the
effective masses of electrons for the in-plane and out-of-plane
motion, respectively. In the case that $H_{c2}$ is a linear function
of temperature, the angular dependence of $\mu_0H_{c2}(\theta)$ can
be converted to that of $T_{c}$ by:\cite{Welp}
\begin{equation}
T_{c}(\theta)=T_{c0}+H/(\partial H_{c2}^{\textbf{H}\parallel
c}/\partial T)\sqrt{cos^2(\theta)+\gamma^{-2}sin^2(\theta)},
\label{eq:4}
\end{equation}
where $T_{c0}$ is the zero field superconducting transition
temperature and \emph{H} is the applied magnetic field. In our case,
the upper critical field near $T_c$ indeed shows nearly linear
temperature dependence for both $\textbf{H}$$\perp$$c$ and
$\textbf{H}$$\parallel$$c$. Therefore, one can estimate the
anisotropic parameter $\gamma$ from the angular dependence of
$T_c(\theta)$. It was shown that a single band anisotropic model can
properly describe the angular dependence of $\mu_0H_{c2}$ in a
multi-band system at temperatures near $T_c$.\cite{ref2,ref1}
Indeed, $T_c(\theta)$ can be nicely fitted by Eq. \ref{eq:4} (see
the solid lines in Fig.\ref{figangle}), indicating that, at least in
the low field region, $T_c(\theta)$ can be described by the G-L
theory and the anisotropic upper critical field is attributed to the
effective mass anisotropy in strongly coupled layered
superconductors. Furthermore, the above fittings give $\gamma$=8.5,
7.3 and 6.3 for $\mu_0H$=3T, 6T and 9T, respectively. These values
of $\gamma(\mu_0H)$ are very close to those shown in the inset of
Fig. 4 if we convert the magnetic fields to temperatures following
the relation of $\mu_0H_{c2}(T_c)$.

In comparison, the iron selenides show intrinsically similar
properties of the upper critical field to the iron pnictide
superconductors, \cite{Yuan,ref2,MHF11,ZJL111,Baily,Kano} but with a
slightly enhanced anisotropy. These findings suggest that all these
FeSCs might share the same characters of superconductivity. This is
surprising because the electronic structure and the normal state of
iron selenides seem to be very unique among the FeSCs. For example,
both hole pockets and electron pockets are observed in iron
pnictides,\cite{arpes} but the hole pocket seems to be absent in
iron selenides.\cite{FDL,DH1,ZXJ} The nesting between the hole
pockets and the electron pockets was regarded as a prerequisite
factor for the forming of $s^{\pm}$ paring state,\cite{Mazin} a
widely accepted proposal for the iron pnictide superconductors. Our
findings of the universal behavior of $\mu_0H_{c2}(T_c)$ in iron
pnictides and selenides, therefore, urge to check whether the
missing of hole pockets in iron selenides is intrinsic or masked by
other experimental factors, e.g., the phase separation or sample
non-stiochiometry. If it is intrinsic, one probably needs to
reconsider the order parameters and the pairing mechanism of FeSCs
in a unified picture.

\section{Thermally Activated Flux Flow}

\begin{figure}
\includegraphics[width=0.45\textwidth]{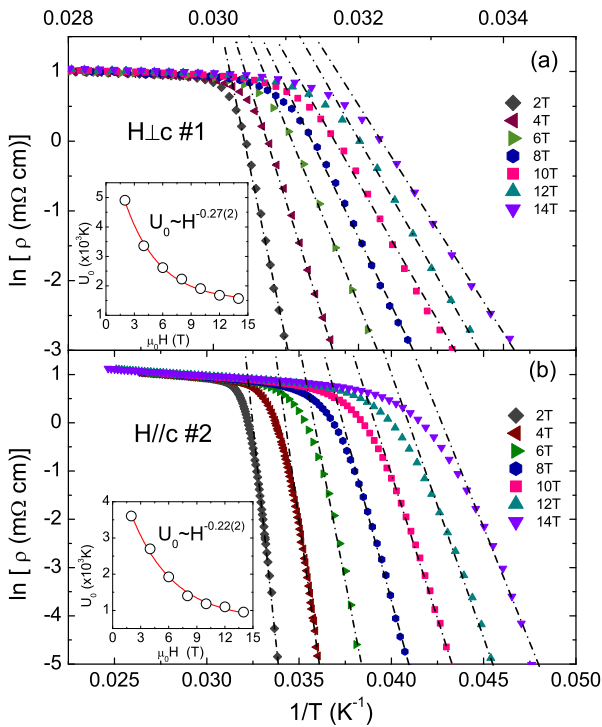}
\caption{(Color online) Arrhenius plot of
Tl$_{0.58}$Rb$_{0.42}$Fe$_{1.72}$Se$_2$ at various magnetic fields:
(a) $\textbf{H}$$\perp$$c$; (b) $\textbf{H}$$\parallel$$c$. The
inset shows the thermally activated energy, $U_{0}(H)$, obtained
from the slope of the Arrehnius plot. The solid lines are fitted to
$U_0(H)\approx H^{-n}$ with n=0.27$\pm$0.02 and 0.22$\pm$0.02 for
$\textbf{H}$$\perp$$c$ and $\textbf{H}$$\parallel$$c$,
respectively.} \label{fig4}
\end{figure}

As already mentioned above, the superconducting transition of
Tl$_{0.58}$Rb$_{0.42}$Fe$_{1.72}$Se$_2$ is significantly broadened
upon applying a magnetic field (see Fig. \ref{fig2DC} and Fig.
\ref{fig2}). Similar features were previously observed in other
layered superconductors, including the cuprates\cite{Kwok,Palstra}
and the 1111-type iron pnictides,\cite{Jaroszynski,Shahbazi,SmFeAsO}
which were interpreted in terms of the energy dissipation caused by
vortices motion. In general, both thermally activated flux flow
(TAFF) and superconducting critical fluctuations may broaden the
resistive superconducting transition in a magnetic field. The
importance of thermal fluctuations is measured by Ginzburg number,
$G_i=\frac{1}{2}$($\gamma$$T_c$/$H_c^2$$\xi_{ab}^3$)$^2$, where
$H_c$ is the thermal dynamic critical field and $\xi_{ab}$ is the
coherence length in the ab-plane. In
Tl$_{0.58}$Rb$_{0.42}$Fe$_{1.72}$Se$_2$, the relatively large
anisotropic parameter $\gamma$ near $T_c$ and the short coherence
length ($\xi\simeq 2.1$ nm) yield a large Ginzburg number $G_i$ and
a soft vortex matter. As a result, thermal fluctuations may become
important enough to overcome the elastic energy of the vortex
lattice in a large part of the magnetic field -temperature phase
diagram, melting the vortex lattice into a liquid. On the other
hand, the critical fluctuations may play a less dominant role here.
Therefore, in the following we will try to interpret the resistive
broadening in terms of the thermally activated flux flows which
actually gives a rather good description to our experimental data.
Further measurements are also under way in order to elucidate the
possible effect of superconducting fluctuations.

According to the TAFF model,\cite{Blatter} the resistivity
$\rho(T,H)$ can be expressed as:
\begin{equation}
\rho(T,H)=(2\nu_0LH/J)\textrm{sinh}[JHVL/T]e^{-J_{c0}HVL/T},
\label{eq:5}
\end{equation}
where $\nu_0$ is an attempt frequency for a flux bundle hopping,
\emph{L} is the hopping distance, \emph{J} is the applied current
density, \emph{$J_{c0}$} is the critical current density in the
absence of flux creep and \emph{V} is the bundle volume. In the
limit of $JHVL/T \ll 1$, Eq. \ref{eq:5} can be simplified as
\begin{equation}
\rho(T,H)=(2\rho_cU/T)e^{-U/T}, \label{eq:6}
\end{equation}
where $U=J_{c0}HVL$ is the thermally activated energy and
$\rho_c$=$\nu_0LH/J_{c0}$. Assuming that $2\rho_cU/T$ is a
temperature independent constant, noted as $\rho_{0f}$, and
$U=U_0(1-T/T_c)$, then Eq.~\ref{eq:6} can be simplified to the
Arrhenius relation:
\begin{equation}
\textrm{ln}\rho(T,H)=\textrm{ln}\rho_{0f}-U_0(H)/T. \label{eq:7}
\end{equation}
Thus, the apparent activation energy, $U_{0}(H)$, could be extracted
from the slopes of the Arrhenius plots, i.e., the plot of ln$\rho$
\emph{v.s.} $1/T$.

In order to study such a possible vortex motion in
Tl$_{0.58}$Rb$_{0.42}$Fe$_{1.72}$Se$_2$, we plot the electrical
resistivity $\rho$ as a function of $1/T$ in a semi-log scale at
various magnetic fields (see Fig.~\ref{fig4}). It is clear that the
Arrhenius relation holds over a wide temperature range for both
$\textbf{H}$$\perp$$c$ and $\textbf{H}$$\parallel$$c$, suggesting
that the TAFF model may nicely describe the field-induced resistive
broadening in Tl$_{0.58}$Rb$_{0.42}$Fe$_{1.72}$Se$_2$. Following
$U_{0}(H)$=-d\textrm{ln}$\rho/\textrm{d}(1/T)$, the apparent
activation energy, $U_0(H)$, can then be determined from the slope
of the linear parts in Fig.~\ref{fig4}. For example, this yields
$U_0$$\approx$4900K for $\textbf{H}$$\perp$$c$ and 3607K for
$\textbf{H}$$\parallel$$c$ at 2T. The derived values of $U_0(H)$ are
plotted as a function of field in the insets of Fig. \ref{fig4}: (a)
$\textbf{H}$$\perp$$c$ and (b) $\textbf{H}$$\parallel$$c$, which
follow a power law of $U_0(H)$$\sim$$H^{-n}$. The fittings give
\emph{n}=0.27$\pm$0.02 for $\textbf{H}$$\perp$$c$ and 0.22$\pm$0.02
for $\textbf{H}$$\parallel$$c$, indicating that the pinning force
may have a weak orientation dependence. The derived values of
$U_0(H)$ are slightly larger than those of some 1111-type iron
pnictides, e.g. NdFeAsO$_{0.7}$F$_{0.3}$\cite{Jaroszynski} and
CeFeAsO$_{0.9}$F$_{0.1}$,\cite{Shahbazi} but smaller than those of
SmFeAsO$_{0.85}$\cite{SmFeAsO} and YBCO\cite{Palstra}, indicating a
moderate pinning force among the cuprates and the FeSCs. Note that
in the 122-type iron pnictides where $\gamma$$\simeq$2 near $T_c$,
the range of vortex liquid state is very narrow and no significant
broadening of the resistive superconducting transition has been
observed in a magnetic field .\cite{Yuan, Maiorov}

\section{Conclusion}

The upper critical field $\mu_0H_{c2}(T_c)$, its anisotropy
$\gamma(T)$ and the vortex motion of the newly discovered
Tl$_{0.58}$Rb$_{0.42}$Fe$_{1.72}$Se$_2$ have been studied by
measuring the field-, temperature- and angular- dependence of the
electrical resistivity. We found that this compound shows a large
upper critical field ($\mu_0H_{c2}^{\textbf{H}\parallel
c}(0\textrm{K})\simeq60$T, $\mu_0H_{c2}^{\textbf{H}\perp
c}(18\textrm{K})\simeq60$T) with a moderate anisotropy near $T_c$.
The anisotropic parameter decreases with decreasing temperature,
reaching $\gamma$(20K)$\sim$2. Analysis based on the WHH model
indicates that the upper critical field is orbitally limited for
$\textbf{H}$$\parallel$$c$, but is likely limited by the spin
paramagnetic effect for $\textbf{H}$$\perp$$c$. These properties
remarkably resemble those of iron pnictide superconductors,
suggesting that all the Fe-based superconductors may bear similar
characters and, therefore, provide restrictions on the theoretical
model. Similar to the cuprates and the 1111-series of iron
pnictides, thermally activated flux flow might be responsible for
the tail structure of the resistive transition below $T_c$ and the
derived thermal activation energies are compatible with those of the
1111-type iron pnictides.

\begin{acknowledgments}
Work at ZJU was supported by the National Science Foundation of
China (grant Nos: 10874146, 10934005), the National Basic Research
Program of China (973 Program) (2009CB929104, 2011CBA00103), the
PCSIRT of the Ministry of Education of China, Zhejiang Provincial
Natural Science Foundation of China, and the Fundamental Research
Funds for the Central Universities. Work at LANL was performed under
the auspices of the National Science Foundation, the Department of
Energy, and the State of Florida. BM was supported by the U.S.
Department of Energy, Basic Energy Sciences, Materials Sciences and
Engineering Division.

\end{acknowledgments}

\end{document}